\documentclass[prd,showpacs,preprint]{revtex4}
\usepackage{graphicx,color,epsfig}

\newcommand{\ba}{\begin{array}}
\newcommand{\ea}{\end{array}}
\newcommand{\bd}{\begin{displaymath}}
\newcommand{\ed}{\end{displaymath}}
\newcommand{\be}{\begin{equation}}
\newcommand{\ee}{\end{equation}}
\newcommand{\beq}{\begin{eqnarray}}
\newcommand{\eeq}{\end{eqnarray}}

\def\deta{\bar\eta}

\newcommand{\non}{\nonumber\\ }

\begin{document}

\title{Determination of  $f_0-\sigma$ mixing angle through \\ $B_s^0 \to J/\Psi~f_0(980)(\sigma)$ decays }

\author{Jing-Wu Li$^{1}${\footnote {lijw@jsnu.edu.cn}},~~~Dong-Sheng Du$^{2}${\footnote {duds@ihep.ac.cn}}~~~Cai-Dian L\"u$^{2}${\footnote {lucd@ihep.ac.cn}}}

\vspace*{1.0cm}

\affiliation{$^{1}$Department of Physics, Jiang Su Normal University,
XuZhou 221116, China,\\$^{2}$Institute of High Energy Physics, P.O.
Box 918(4), Beijing 100049, China}


\vspace*{1.0cm}

\date{}
\begin{abstract}
We study $B_s^0 \to J/\psi f_0(980)$ decays, the quark content of $f_0(980)$ and the mixing angle of $f_0(980)$ and $\sigma(600)$. We calculate not only the factorizable contribution in QCD
facorization scheme but also the nonfactorizable hard spectator corrections in QCDF and pQCD approach. We get consistent result with the
experimental data of $B_s^0 \to J/\psi f_0(980)$ and predict the branching ratio of $B_s^0 \to J/\psi \sigma$. We suggest two ways to determine
$f_0-\sigma$ mixing angle $\theta$. Using the experimental measured branching ratio of $B_s^0
\to J/\psi f_0(980) $, we can get the $f_0-\sigma$ mixing angle
$\theta$ with some theoretical uncertainties. We suggest another way to
determine $f_0-\sigma$ mixing angle $\theta$ using both of
experimental measured decay branching ratios $B_s^0 \to J/\psi
f_0(980) (\sigma)$  to avoid theoretical uncertainties.

\end{abstract}

 \maketitle


\section{Introduction}


Scalar mesons are important for testing QCD and the Standard Model(SM).
 Many scalar mesons  have been observed: isoscalar states
$\sigma(600)$, $f_0(980)$, $f_0(1500)$, $f_0(1370)$, $f_0(1710)$; the isovector states $ a_0(980)$, $a_0(1450)$ and isodoublets $\kappa(800)$,
$K^*_0(1430)$ \cite{pdg2010}. The number of these scalar mesons exceeds the particle states which can be accommodated in one nonet in the
quark model. It is commonly believed that there are two nonets below and above 1GeV \cite{close1}-\cite{scalar}. The  meson states in each nonet
have not been completely determined yet. Especially,the structure of $f_0(980)$ (abbreviated as $f_0$)  is not settled.  The underlying
structure of $f_0(980)$ concerns the extraction of the  $CP$-violating phase
   $ \beta_s$ in $B^0_s-\bar B^0_s$ mixing, defined as $ \beta_s= Arg\left[ - \frac{V_{ts}
V^*_{tb}}{V_{cs} V^*_{cb}}\right]$, which is particularly important
 for look for  new physics(NP)\cite{betas1}-\cite{betas4}.  The $CP$-violating phase
   $ \beta_s$  is predicted to be tiny in the
SM: $\beta_s \simeq 0.019$ rad.  This is about 20 times smaller in magnitude than the measured value of the corresponding phase $2\beta$ in
${B}^0-\bar {B}^0$ mixing. Being small, this phase can be drastically increased by the presence of new physics beyond the SM.  Thanks to the
suppression of light-quark loops, $ \beta_s$ is dominated by short-distance processes and sensitive to NP. Thus, measuring $\beta_s$ is an
important probe for new physics.

Attempts to determine $\beta_s$ have been made by the CDF, D0, LHCb and ATLAS Collaboration   based on  the angular analysis of $B_s\to J/\psi
\phi$\cite{Tevatron0,2008fj,11123183,Aaltonen:2007he}. Reliable signal of new physics is not founded based on the measured datas of $\beta_s$, because of  sizable
uncertainties due to the strong phases involved in
 the angular analysis of $B_s\to J/\psi \phi$\cite{dunietzpball}. So  the precise measurement of $\beta_s$ is one of the priorities in the physics programs at the hadron colliders and at the $B$
factories  \cite{betas4,machines}. In Ref.~\cite{betas1} it is argued that in the case of $J/\psi\phi$ final
 state the analysis is complicated by the presence of an S-wave $K^+K^-$ system interfering with the $\phi$.  So it is necessary to consider other
process to access mixing parameter $\beta_s$.

$B_s^0\to J/\psi f_0(980)$, which has been observed by the LHCb ,
 CDF and Belle Collaboration recently\cite{betas2,betas3,11104272},
  is another promising channel for
accessing the mixing parameter. The advantage of this channel is clear: no angular analysis is required because of the $J^P=0^+$ quantum numbers
of the $f_0(980)$. To determine the phase $\beta_s$ through $B_s^0\to J/\psi f_0(980)$, it is essential to study the structure of $f_0(980)$.

The structure of $f_0(980)$ have been investigated in many works\cite{f09801}-\cite{f09803}. Studies show that $f_0(980)$ is not a pure $s\bar
s$ state. The  First experimental evidence is  the observation of $\Gamma(J/\psi\to f_0\omega)\approx {1\over 2}\Gamma(J/\psi\to f_0\phi)$. This
result clearly indicates the existence of both the non-strange and strange quark content in $f_0(980)$. The Second evidence  is that $f_0(980)$
and $a_0(980)$ have similar widths and that the $f_0$ width is dominated by $\pi\pi$, that means  the existence of $u\bar u$ and $d\bar d$ pairs
in $f_0(980)$. So, $f_0(980)\to\pi\pi$ should not be OZI suppressed relative to $a_0(980)\to\pi\eta$. Therefore, isoscalars $\sigma(600)$ and
$f_0$ must have  mixing\cite{chy0508104},
\begin{eqnarray}
 |f_0(980)\rangle = |s\bar s\rangle\cos\theta+|n\bar n\rangle\sin\theta,
 \qquad |\sigma(600)\rangle = -|s\bar s\rangle\sin\theta+|n\bar n\rangle\cos\theta,
 \end{eqnarray}
with $n\bar n\equiv (\bar uu+\bar dd)/\sqrt{2}$ and $\theta$ is  $f_0\!-\!\sigma$ mixing angle.

Many attempts have been made to determine the $f_0\!-\!\sigma$ mixing angle. Analysis of experimental data shows that the $f_0\!-\!\sigma$
mixing angle $\theta$ lies in the ranges of $25^\circ<\theta<40^\circ$ and $140^\circ<\theta< 165^\circ$\cite{f09801}-\cite{f09802}. The
$f_0\!-\!\sigma$ mixing angle is generally determined
 through the calculation of branching ratios of some mesons decays.
 In the calculation of  the mesons decay amplitudes, some parameters have to be taken as inputs, so the determination of $f_0\!-\!\sigma$ mixing angle
 has many uncertainty sources, such as, decay constant, transition form factors,
hadron  coupling constants, wave functions of the relevant mesons, and assumptions about the variation of the form factors with momentum
transfer $Q^2$.  It is not a good way to determine $f_0\!-\!\sigma$ mixing angle with too many parameters and assumptions. To extract $\beta_s$
with better accuracy, it is necessary to  find a better method to determine it with less input parameters.

Based on only one conventional  assumption that the decay constant and the distribution amplitude  of the $s\bar{s}$ component for $f_0$
  is the same as that for  $\sigma$ as in Ref.~(\cite{feldmann1}-\cite{ball}),
  we can derive the relation between
  the branching ratios of $B_s^0\to J/\Psi~f_0(\sigma)$ in Eq.~(\ref{rbrf0withsigma2}).
From this relation, we can determine $f_0\!-\!\sigma$ mixing angle.   The only input we need is the experimental value of the ratio of the
branching ratios for $B_s^0\to J/\Psi~f_0(\sigma)$. That means that the  $f_0\!-\!\sigma$ mixing angle determined in this way has much less
uncertainty sources.

This paper is organized as follows. In Sec.~2, we derive the formulas for the amplitudes of the $B_s^0\to J/\Psi~f_0(\sigma)$. Two methods for
determining the $f_0\!-\!\sigma$ mixing angle are presented. Section 3 is for summary and discussion. Some input parameters and mesons wave
function are listed in the Appendix.

\section{Branching Ratios for the Decays of $B_s^0 \to J/\psi f_0(980)$}
\label{br}

For the $B_s^0 \to J/\psi f_0(980)$ decays, the effective
Hamiltonian is given by \cite{buras96},
\begin{eqnarray}
H_{\rm eff}&=&\frac{G_F}{\sqrt
2}\left\{V_{cb}^{*}V_{cs}[C_1(\mu)O_1+
C_2(\mu)O_2]-V_{tb}^{*}V_{ts}\sum_{k=3}^{10}C_k(\mu)O_k\right\}\;,
\label{effh}
\end{eqnarray}
with the Cabibbo-Kobayashi-Maskawa (CKM) matrix elements $V$ and the four-fermion operators,
\begin{eqnarray}
& &O_1 = (\bar{b}_ic_j)_{V-A}(\bar{c}_j s_i)_{V-A}\;,\;\;\;\;\;\;\;\;
O_2 = (\bar{b}_ic_i)_{V-A}(\bar{c}_j s_j)_{V-A}\;,
\nonumber \\
& &O_3 =(\bar{b}_i s_i)_{V-A}\sum_{q}(\bar{q}_j q_j)_{V-A}\;,\;\;\;\;
O_4 =(\bar{b}_i s_j)_{V-A}\sum_{q}(\bar{q}_jq_i)_{V-A}\;,
\nonumber \\
& &O_5
=(\bar{b}_is_i)_{V-A}\sum_{q}(\bar{q}_jq_j)_{V+A}\;,\;\;\;\;O_6
=(\bar{b}_is_j)_{V-A}\sum_{q}(\bar{q}_jq_i)_{V+A}\;,
\nonumber \\
& &O_7
=\frac{3}{2}(\bar{b}_is_i)_{V-A}\sum_{q}e_q(\bar{q}_jq_j)_{V+A}\;,
\;\; O_8
=\frac{3}{2}(\bar{b}_is_j)_{V-A}\sum_{q}e_q(\bar{q}_jq_i)_{V+A}\;,
\nonumber \\
& &O_9
=\frac{3}{2}(\bar{b}_is_i)_{V-A}\sum_{q}e_q(\bar{q}_jq_j)_{V-A}\;,
\;\; O_{10}
=\frac{3}{2}(\bar{b}_is_j)_{V-A}\sum_{q}e_q(\bar{q}_jq_i)_{V-A}\;,
\end{eqnarray}
$i, \ j$ being the color indices.

In this paper, we take the light-cone coordinates $(p^+, p^-, {\bf
p}_T)$ to describe the four-dimensional momenta of the meson,
\begin{eqnarray} p^\pm = \frac{1}{\sqrt{2}} (p^0 \pm p^3), \quad and \quad
{\bf p}_T = (p^1, p^2).
 \end{eqnarray}

At the rest frame of the $B_s^0$ meson, the momentum $P_{1}$ of the
$B_s^0$ meson is
\begin{eqnarray}
  P_1 &=& \frac{M_{B_s}}{\sqrt{2}} (1,1,{\bf 0}_T)
\end{eqnarray}

the $J/\Psi (f_0)$ meson momentum $P_{2}(P_{3})$ can be written as

\begin{eqnarray}
  \quad P_2 =
\frac{M_{B_s}}{\sqrt{2}}(1-r_3^2,r_2^2,{\bf 0}_T), \quad P_3 =
\frac{M_{B_s}}{\sqrt{2}} (r_3^2,1-r_2^2,{\bf 0}_T)
\end{eqnarray}

with $r_2=m_{J/\psi}/M_{B_s}$, $r_3=m_{f_0}/M_{B_s}$.

The polarization  vectors of the $J/f_0$ meson are parameterized as
\begin{eqnarray}
\epsilon_{2L}=\frac{1}{\sqrt{2}r_2}\left(1,-r_2^2, {\bf
0}_T\right)\;,\;\;\;\; \epsilon_{2T}=\left(0,0, {\bf
1}_T\right)\;.\label{pol}
\end{eqnarray}

The decay width of of $B_s^0 \to J/\psi f_0(980)$ is

\begin{equation}\label{gamma}
\Gamma=\frac{1}{32\pi M_{B_s}}G_F^2(1-r_2^2+\frac{1}{2}r_{2}^4-r_3^2)
|{\cal A}|^2\;.
\end{equation}
  The amplitude ${\cal A}$ consists of  factorizable part and
  nonfactorizable part.  It can be written as
\begin{eqnarray}
{\cal A}&=&A_{FA}+ A_{VERT}+ A_{HS}\;, \label{ampeta}
\end{eqnarray}

where $A_{FA}$ denotes the factorizable contribution, $A_{VERT}$ is
the vertex corrections from Fig.~\ref{nonfc}.(a)-(d), $A_{HS}$ is
the hard spectator scattering correction from
Fig.~\ref{nonfc}.(e)-(f).

\subsection{Factorizable Contribution and Vertex Correction In QCDF}
\begin{figure}[htb]
\begin{center}
\epsfig{file=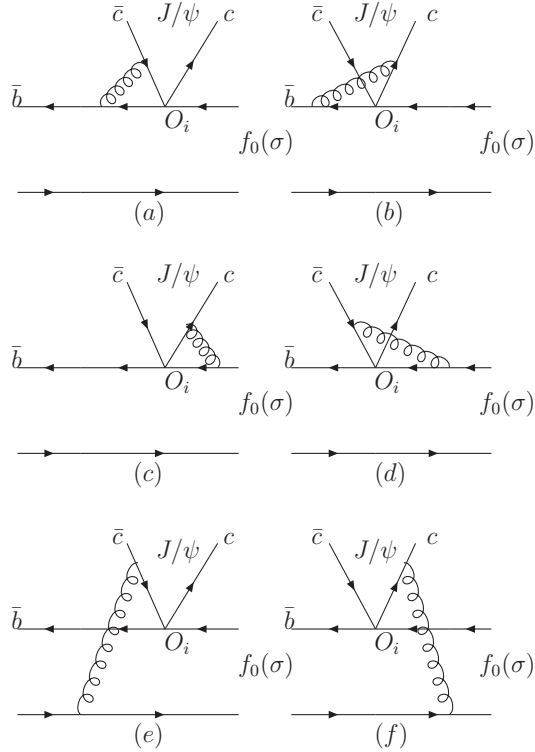,height=10cm,width=7cm,angle=0}
\caption{ Nonfactorizable contribution to the $B_s^0\to
J/\Psi~f_0(\sigma)$ decays} \label{nonfc}
\end{center}
\end{figure}

The factorizable part $A_{FA}$ of amplitude ${\cal A}$ in
Eq.~(\ref{ampeta}) for $B_s \to J/\psi f_0(980)$ decay  can not be
 calculated reliably in pQCD approach, because its characteristic scale is around 1 GeV \cite{cljpsi}.
 We here compute the factorizable part of amplitude and the vertex correction
from Fig.~\ref{nonfc}.(a)-(d) in QCDF \cite{BBNS} instead of pQCD
approach and get

\begin{eqnarray}
A_{FA}+A_{VERT}=a_{eff} m_{B_s}^2 \cos\theta f_{J/{\psi}}F_1^{B_s\to
f_0}(m_{J/\psi}^2) (1-r_2^2)\;,\label{amppinf}
\end{eqnarray}
where $f_{J/{\psi}}$ is decay constant of $J/\psi$ meson, $F_1^{B_s\to
 f_0}$ is the $B_s \to  f_0$ transition form
factor defined as
\begin{eqnarray}
&&\langle f_0(P_3)|{\bar b}\gamma_\mu \gamma_5 s|B_s(P_1)\rangle =\nonumber\\&&-i \Big\{F_1^{B_s\to
f_0}(q^2)[(P_1+P_3)_\mu-\frac{m_{B_s}^2-m_{f_0}^2}{q^2}q_\mu]
+F_0^{B_s\to f_0}(q^2)\frac{m_{B_s}^2-m_{f_0}^2}{q^2}q_\mu\Big\},\label{fp}
\end{eqnarray}

$q=P_1-P_3$ being the momentum transfer, and $m_{f_0}$ the $f_0$
meson mass.

The Wilson coefficient $a_{eff}$ for $B_s \to J/\psi f_0(980)$ can
be derived in QCDF\cite{BBNS2,qcdfvc},
\begin{eqnarray}
a_{eff}&=&V_c^\ast \left[C_1+
\frac{C_2}{N_c}+\frac{\alpha_s}{4\pi}\frac{C_F}{N_c}C_2
\left(-18+12\ln\frac{m_b}{\mu}+f_I\right)\right]\nonumber\\&&-V_t^\ast
\Big[C_3+\frac{C_4}{N_c}+\frac{\alpha_s}{4\pi}\frac{C_F}{N_c}C_4
\left(-18+12\ln\frac{m_b}{\mu}+f_I\right)\nonumber\\&&+C_5+\frac{C_6}{N_c}+\frac{\alpha_s}{4\pi}\frac{C_F}{N_c}C_6
\left(6-12\ln\frac{m_b}{\mu}-f_I\right)\nonumber\\&&+C_7+\frac{C_8}{N_c}+\frac{\alpha_s}{4\pi}\frac{C_F}{N_c}C_8
\left(6-12\ln\frac{m_b}{\mu}-f_I\right)\nonumber\\&&+C_9+\frac{C_{10}}{N_c}+\frac{\alpha_s}{4\pi}\frac{C_F}{N_c}C_{10}
\left(-18+12\ln\frac{m_b}{\mu}+f_I\right)\Big]\;,\nonumber\\
\label{anfandver}
\end{eqnarray}

with the function,
\begin{eqnarray}
f_I=\frac{2\sqrt{2N_c}}{f_{J/\psi}}\int
dx_2\Psi^L(x_2)\left[\frac{3(1-2x_2)}{1-x_2}\ln x_2-3\pi
i+3\ln(1-r_2^2)+\frac{2r_2^2(1-x_2)}{1-r_2^2 x_2}\right]\;,
\end{eqnarray}

and $V_c^\ast=V_{cb}^{*}V_{cs}$, $V_t^\ast=V_{tb}^{*}V_{ts}$.

For the $B_s\to f_0$ transition form factor, we employ the models
derived from the light-cone sum rules \cite{betas4}, which is
parameterized as
\begin{eqnarray}
F_1^{B_s\to
f_0}(q^2)=\frac{F_1^{B_s\to
f_0}(0)}{1-a_1q^2/m_{B_s}1^2+b_1q^4/m_{B_s}^4}\;\label{f1bstof0}
\end{eqnarray}

with $F_1^{B_s\to
f_0}(0)=0.238$, $a_1=1.5$, $b_1=0.58$, for $B_s \to f_0$
transition.

\subsection{Hard Spectator Scattering Corrections In QCDF Approach}

For the contribution $A_{HS}$ from hard spectator scattering corrections in
Fig.~\ref{nonfc}.(e)-(f), we can use QCD factorization and get,
\begin{eqnarray}
A_{HS}&=&
 \cos\theta f_{J/{\psi}}{C_F\alpha_s \pi\over N_c^2}(V_c^\ast C_2+ V_t^\ast (2C_6+2C_8-C_4-C_{10})) H_1(M_1M_2)\nonumber \;,\label{eq:amppinf1}
\end{eqnarray}

where $H_{1}(M_1M_2)$ is the hard spectator function,
 \begin{eqnarray} \label{eq:H1}
 H_1(M_1M_2) &=& -f_{B_s}\bar{f}_{f_0}\int^1_0 {d\rho\over\rho}\, \Phi_{B_s}(\rho)\int^1_0 {d\xi\over
\bar\xi} \,\Phi_{J/{\psi}}(\xi)\int^1_0 {d\eta\over \deta}\left[-
\Phi_{f_0}(\eta)+ r_\chi^{f_0}\,{\bar\xi\over
\xi}\,\Phi_{f_0}^s(\eta)\right], \nonumber \\
 \end{eqnarray}

 Because twist-3 distribution amplitude $\Phi_{f_0}^s$ of $f_0$ meson  is
 $\Phi_{f_0}^s=\bar {f}_{f_0}$\cite{chy07053079}, the integral
 $\int^1_0 {d\eta\over \deta}\Phi_{f_0}^s(\eta)$ will generate logarithmical divergence from end-point. It is often  parameterized  as\cite{BBNS2},
 \begin{equation}\label{eq:divpara}
 \int^1_0 {d\bar{\eta}\over \deta}= \ln(\frac{m_{B_s}}{\Lambda_{QCD}})+ r\exp(i\delta)
 \end{equation}
  where parameter $r $ is often taken  from 0 to 6, $\delta$ is phase, $0 \leq\delta\leq 2\pi$.

\begin{figure}[htb]
\begin{center}
\psfig{file=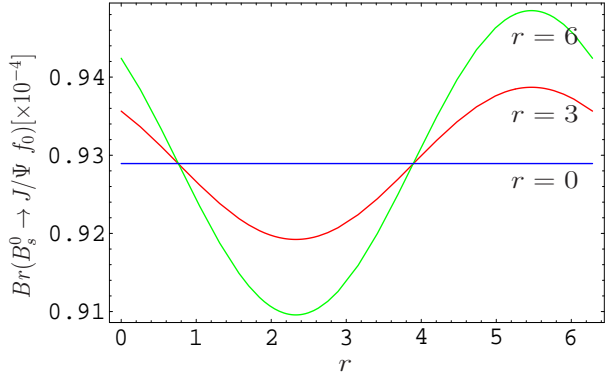,height=5cm,width=8cm}
\end{center}
 \caption{The variation of the
 the branching ratios of $B_s^0\to J/\Psi~f_0$ with phase $\delta$, three curves are  for parameter $r=0,3,6$, respectively. }\label{fig:reqvary}
\end{figure}

 According to Eq.~(\ref{gamma})-(\ref{eq:divpara}), taking the parameter $r$ varying from 0 to 6, $0 \leq\delta\leq 2\pi$ and other parameters listed in the Appendix, we can get the branching ratio of $B_s \to J/\psi f_0(980)$. The range of predicted the branching ratio of $B_s \to J/\psi f_0(980)$ is,
\begin{figure}[htb]
\begin{center}
\psfig{file=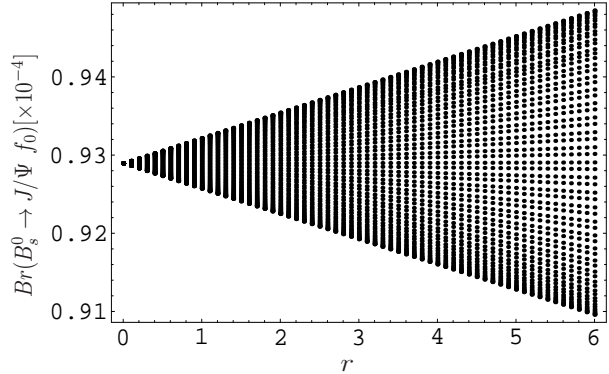,height=5cm,width=8cm}
\end{center}
 \caption{The range  of  the branching ratios of $B_s^0\to J/\Psi~f_0$, parameter r varies from 0 to 6 and $\delta$ from 0 to $2\pi$}\label{fig:parainf}
\end{figure}

 \begin{equation}
 9.1 \times 10^{-5} < Br(B_s \to J/\psi f_0(980))< 9.6\times 10^{-5},
 \end{equation}

 which is shown from Fig.~(\ref{fig:parainf}),(\ref{fig:reqvary}).

 With the value of the branching ratio of $f_0(980)\to\pi^+\pi^-$ in
 Ref.~\cite{chy07053079},
 \begin{equation}\label{chyf0topipi}
 Br(f_0(980)\to \pi^+\pi^-)=0.50^{+0.07}_{-0.09}
 \end{equation}

we can get the branching ratio of
$B_s^0\to J/\psi f_0(980);f_0(980)\to\pi^+\pi^-$,
\begin{equation}\label{brbstojpsif02}
   4.55 \times 10^{-5} <Br(B_s ^0\to J/\Psi~f_0; f_0(980)\to\pi^+\pi^-)< 4.8\times 10^{-5}
\end{equation}
 This prediction of the branching ratio is about half of  the averaged experimental data\cite{betas2,betas3,11104272} ,
\begin{equation}\label{belldata}
    Br^{exp}(B_s ^0\to J/\Psi~f_0;f_0(980)\to\pi^+\pi^-)=(1.20^{+0.25}_{-0.21}(\mathrm{stat.})^{+0.17}_{-0.19}(\mathrm{syst.})
) \times 10^{-4}
\end{equation}

 So, it seems that the result from QCDF can not accomodate the experimental data.  The  reason is that the  divergent integral
 in hard spectator correction is approximately expressed by the parameters, which are  suitable for the modes in which hard spectator correction
 has little contribution. .
\subsection{Hard Spectator Scattering Corrections In pQCD Approach}

  The divergence in the hard spectator correction arises from the neglect of transverse
 momentum. Using pQCD approach can avoid the divergence in the calculation of  the  hard spectator scattering corrections
  because transverse momentum of quarks is kept. The characteristic hard scale
in the hard spectator scattering corrections is  higher than that in $B_s$ meson transition form factor~\cite{Chou:2001bn}. Therefore, we can
employ pQCD approach based on $k_T$ factorization theorem, which is free from the end-point singularity for the spectator amplitude
\cite{cljpsi}. In pQCD approach,the nonfactorizable hard spectator amplitudes can be written as,
\begin{eqnarray}
A_{HS}&=&V_c^\ast {\cal M}_{1}^{(J/\psi f_0)}- V_t^\ast {\cal
M}_{4}^{(J/\psi f_0)}-V_t^\ast{\cal M}_{6}^{(J/\psi
f_0)}\;,\label{amppihs}
\end{eqnarray}
where the amplitudes ${\cal M}_{1,4}^{(J/\psi f_0)}$ and ${\cal M}_{6}^{(J/\psi f_0)}$ come from the $(V-A)(V-A)$ and $(V-A)(V+A)$ operators
in Eq.~(\ref{effh}), respectively. Their factorization
formulas are given by pQCD approach

\begin{eqnarray}
{\cal M}_{1,4}^{(J/\psi f_0)} &=&8\pi m_{B_s}^4 C_{F}\sqrt{2N_{c}}%
\int_{0}^{1}[dx]\int_{0}^{\infty }b_{1}db_{1}b_{2}db_{2} \Phi _{B_s}(x_{1},b_{1})
\nonumber \\
&& \times \Big\{ \Big[ ((r_2^2 - 1)(\psi^t(x_2)r_2^2 + 2\psi^L(x_2)(r_2^2 - 1)(x_1 + x_2 - 1))\phi_{f0}(x_3)\nonumber \\
&& +
   2\psi^L(x_2)r_3(((x_1 - x_3)r_2^2 + x_3)\phi_{f_0}^s(x_3)\nonumber \\
&& +
     (x_3 - r_2^2(x_1 + 2x_2 + x_3 - 2))\phi_{f_0}^{\sigma}(x_3)))
  \Big]\nonumber \\
&&\times E_{1,4}(t_d^{(1)})h_d^{(1)}(x_1,x_2,x_3,b_1)
\nonumber \\%
&& - \Big[ (r_2^2 - 1)(\psi^t(x_2)r_2^2 - 2\psi^L(x_2)(x_2r_2^2 - x_3r_2^2 - x_1 + x_2 + x_3))\phi_{f_0}
   x_3\nonumber \\
&& - 2r_3(\psi^L(x_2)((x_1 - x_3)r_2^2 + x_3)\phi_{f_0}^s(x_3)\nonumber \\
&& +
    (2\psi^t(x_2)r_2^2 + \psi^L(x_2)(r_2^2(x_1 - 2x_2 + x_3) - x_3))\phi_{f_0}^{\sigma}(x_3))\Big]
 \nonumber \\
&& \times E_{1,4}(t^{(2)}_d)
h_d^{(2)}(x_1,x_2,x_3,b_1)\;,\label{eq:m1,4}
\end{eqnarray}

\begin{eqnarray}
{\cal M}_{6}^{(J/\psi f_0)} &=&8\pi m_{B_s}^4 C_{F}\sqrt{2N_{c}}%
\int_{0}^{1}[dx]\int_{0}^{\infty }b_{1}db_{1}b_{2}db_{2} \Phi _{B_s}(x_{1},b_{1})
\nonumber \\
&& \times \Big\{  \Big[
(r_2^2 - 1)(\psi^t(x_2)r_2^2 + 2\psi^L(x_2)((x_2 + x_3 - 1)r_2^2 + x_1 + x_2 - x_3 - 1))
   \phi_{f_0}x_3\nonumber \\
&& - 2 r3(\psi^L(x_2)((x1 - x3)r_2^2 + x_3)\phi_{f_0}^s x_3 \nonumber \\
&&+
    (2\psi^t(x_2)r2^2 + \psi^L(x_2)(r2^2(x_1 + 2x_2 + x_3 - 2) - x_3))\phi_{f_0}^{\sigma}(x3))\Big]\nonumber \\
&&\times E_{6}(t_d^{(1)})h_d^{(1)}(x_1,x_2,x_3,b_1)
\nonumber \\%
&& - \Big[ (r_2^2 - 1)(\psi^t(x_2)r_2^2 + 2\psi^L(x_2)(r_2^2 - 1)(x_1 - x_2))\phi_{f_0}x_3\nonumber \\
&& +
  2\psi^L(x_2)r3(((x_1 - x_3)r_2^2 + x_3)\phi_{f_0}^s x_3 \nonumber \\
&&+
    (x_3 - r_2^2(x_1 - 2x_2 + x_3))\phi_{f_0}^{\sigma}x_3)
  \Big]\nonumber \\
&& \times E_{6}(t^{(2)}_d) h_d^{(2)}(x_1,x_2,x_3,b_1)
\Big\}\;,\label{psi6}
\end{eqnarray}
with the color factor $C_F=4/3$, the number of colors $N_c=3$, the
symbol $[dx]\equiv dx_1 dx_2 dx_3$ and the mass ratio
$r_{f_0}=m_0^{f_0^{\bar{s}s}}/m_{B_s}$, $m_0^{f_0^{\bar{s}s}}$ being the
chiral scale associated with the $f_0$ meson. In the
calculation of ${\cal M}_{1,4}^{(J/\psi f_0)}$ and ${\cal
M}_{6}^{(J/\psi f_0)}$,  we reserve  the power terms of $r_2$ up to
$\mathcal{O}(r^{4}_{2})$, the power terms of $r_3$ up to
$\mathcal{O}(r^{2}_{3})$, because  $J/\psi$ meson is heavy.

In the derivation of spectator correction in pQCD approach, we need
to input the wave function of relevant mesons, we list the wave
functions in appendix.The evolution factors are written as\cite{cljpsi}
\begin{eqnarray}
E_{i}(t) &=&\alpha _{s}(t) a_{i}^{\prime}(t)S(t)|_{b_{3}=b_{1}}\;,
\end{eqnarray}
with the Wilson coefficients,
\begin{eqnarray}
a_{1}^{\prime } &=&\frac{C_{2}}{N_{c}};, \nonumber\\
a_{4}^{\prime } &=&\frac{1}{N_{c}} \left(
C_{4}+\frac{3}{2}e_{c}C_{10}\right)\;, \nonumber\\
a_{6}^{\prime} &=&\frac{1}{N_{c}}\left(
C_{6}+\frac{3}{2}e_{c}C_{8}\right)\;.
\end{eqnarray}
The Sudakov exponent is given by\cite{cljpsi}
\begin{eqnarray}
S(t)&=&S_{B_s}(t)+S_{f_0}(t)\;,\nonumber\\
S_{B_s}(t)&=&\exp\left[-s(x_{1}P_{1}^{+},b_{1})
-\frac{5}{3}\int_{1/b_{1}}^{t}\frac{d{\bar{\mu}}} {\bar{\mu}} \gamma
(\alpha _{s}({\bar{\mu}}))\right]\;,
\label{sb} \nonumber \\
S_{f_0}(t)&=&\exp\left[-s(x_{3}P_{3}^{-},b_{3})
-s((1-x_{3})P_{3}^{-},b_{3})
-2\int_{1/b_{3}}^{t}\frac{d{\bar{\mu}}}{\bar{\mu}} \gamma
(\alpha_{s}({\bar{\mu}}))\right]\;, \label{sbk}
\end{eqnarray}

The hard functions $h_d^{(j)}$, $j=1$ and 2, are
\begin{eqnarray}
h^{(j)}_d&=& \left[\theta(b_1-b_2)K_0\left(DM_{B_s}
b_1\right)I_0\left(DM_{B_s}b_2\right)\right. \nonumber \\
& &\quad \left.
+\theta(b_2-b_1)K_0\left(DM_{B_s} b_2\right)
I_0\left(DM_{B_s} b_1\right)\right]
\nonumber \\
&  & \times  K_{0}(D_{j}M_{B_s}b_{2})\;,\;\;\;\;\;\;\;\;\;\;\;\;\;\;
\mbox{for $D^2_{j} \geq 0$}\;,
\nonumber  \\
&  & \times \frac{i\pi}{2} H_{0}^{(1)}(\sqrt{|D_{j}^2|}M_{B_s}b_{2})\;,\;\;\;\;
 \mbox{for $D^2_{j} \leq 0$}\;,
\label{hdf}
\end{eqnarray}
with the variables,
\begin{eqnarray}
D& =&  x_1x_3 - x_1x_3r_2^2 - r_3^2x_3^2 \;, \nonumber \\
D_1& =& x_1x_3 + x_2x_3 - x_3+(-x_2^2 - x_1x_2 - x_3x_2 + 2x_2 + x_1
- x_1x_3 + x_3 - 1)r_2^2 \nonumber\\&&
  + r_3^2(-x_3^2 - x_2x_3 + x_3)+\frac{1}{4}r_2^{2} \;,\nonumber\\
D_2& =&  x_1x_3 - x_2x_3 +(-x_2^2 + x_1x_2 + x_3x_2 - x_1x_3)r_2^2 +
  r_3^2(x_2x_3 - x_3^2)+\frac{1}{4}r_2^{2}
\;.
\end{eqnarray}
In the calculation of hard function, Considering the heavy $J/\psi$ meson,  we  reserve the power terms of $r_2$ up  to $\mathcal{O}(r^{4}_{2})$, the power terms of $r_3$ up to
$\mathcal{O}(r^{2}_{3})$.

 The hard scales $t$ are chosen as
\begin{eqnarray}
t^{(j)}={\rm max}(\sqrt{D}m_{B_s},\sqrt{|D_j|}m_{B_s},1/b_1)\;.
\end{eqnarray}

\subsection{Numerical Analysis}
From the Eq.~(\ref{gamma}) and Eq.~(\ref{ampeta}), we can derive the
relation of the  branching ratio of $B_s \to J/\psi f_0(\sigma)$ with
 $f_0-\sigma$ mixing
angle $\theta$,

\begin{eqnarray}\label{brbstojpsif0}
  Br(B_s^0\to J/\Psi~f_0) &=& \frac{1}{32\pi M_{B_s} \Gamma_{B_s^0}}G_F^2(1-r_2^2+\frac{1}{2}r_{2}^4-r_{3(f_0)}^2)\nonumber\\
                         &&\cos^2{\theta}
  |(A_{FA}+ A_{VERT}+ A_{HS})|^2
\end{eqnarray}

\begin{eqnarray}\label{brbstojpsisigma}
  Br(B_s^0\to J/\Psi~\sigma) &=& \frac{1}{32\pi M_{B_s} \Gamma_{B_s^0}}G_F^2(1-r_2^2+\frac{1}{2}r_{2}^4-r_{3(\sigma)}^2)\nonumber\\
                         &&\sin^2{\theta}
  |(A_{FA}+ A_{VERT}+ A_{HS})|^2
\end{eqnarray}
where $r_{3(f_0)}=m_{f_0}/m_{B_s}$,
$r_{3(\sigma)}=m_{\sigma}/m_{B_s}$, $\Gamma_{B_s^0}$ is the total decay width
of $B_s^0$ meson.

 To calculate the the branching ratio of
$B_s^0\to J/\Psi~f_0$, it is  necessary to take some parameters and distribution amplitude for relevant mesons as inputs.

 The parameters and
distribution amplitude for the relevant mesons used in this paper are listed in the Appendix.

   The  $f_0-\sigma$ mixing angle $\theta$ lies in the ranges $25^\circ<\theta<40^\circ$
and $140^\circ<\theta< 165^\circ$\cite{chy0508104}.
According to Eq.~(\ref{brbstojpsif0}), we can get the branching ratio of
$B_s^0\to J/\psi~f_0$,
\begin{equation}
    Br(B_s ^0\to J/\psi~f_0)= (2.43 ^{+0.30}_{-0.31}(\omega_{B_s})) \times 10^{-4},
\end{equation}
The main theoretical error of $Br(B_s ^0\to J/\psi~f_0)$ is   induced by
 the uncertainty of shape factor $\omega_{B_s}$ of $B_s$ meson wave function in Eq.~(\ref{eq:phibs}).

    With the value of the branching ratio of $f_0(980)\to\pi^+\pi^-$ in
in Eq.~(\ref{chyf0topipi}),
 we can get the branching ratio of
$B_s^0\to J/\psi f_0(980);f_0(980)\to\pi^+\pi^-$,
\begin{equation}\label{brbstojpsif02}
    Br(B_s ^0\to J/\Psi~f_0; f_0(980)\to\pi^+\pi^-)=(1.215 ^{+0.15}_{-0.155}(\omega_{B_s})^{+0.17}_{-0.21}(f_0))\times 10^{-4}
\end{equation}

The first theoretical error of $Br(B_s ^0\to J/\Psi~f_0; f_0(980)\to\pi^+\pi^-)$  is from the uncertainty of shape factor $\omega_{B_s}$ of $B_s$ meson wave function, the second  one is induced by the uncertainty of
 the branching ratio of $f_0(980)\to\pi^+\pi^-$ in Eq.~(\ref{chyf0topipi}).

Compared with the averaged experimental data\cite{betas2,betas3,11104272} ,
\begin{equation}\label{belldata}
    Br^{exp}(B_s ^0\to J/\Psi~f_0;f_0(980)\to\pi^+\pi^-)=(1.20^{+0.25}_{-0.21}(\mathrm{stat.})^{+0.17}_{-0.19}(\mathrm{syst.})
) \times 10^{-4}
\end{equation}
    our prediction is in consistency with  the experimental
   value.

\begin{figure}[htb]
\begin{center}
\psfig{file=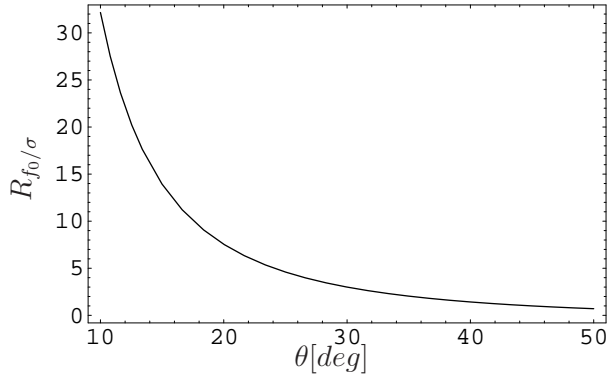,height=5cm,width=8cm}
\end{center}
 \caption{The
ratio of the branching ratios of $B_s^0\to J/\Psi~f_0$ to that of
$B_s^0\to J/\Psi~\sigma$ as function of $f_0-\sigma$ mixing angle
$\theta$ }\label{vrf0sigma}
\end{figure}

There are two ways to extract the  $f_0-\sigma$ mixing
angle $\theta$. One is
   based on the relation of the  branching ratio of $B_s \to J/\psi f_0(980)$ with
 mixing angle $\theta$ shown in Eq.(\ref{brbstojpsif0}). We can  determine $f_0-\sigma$ mixing
angle $\theta$.

\begin{equation}\label{theta1}
 \theta=(34.03^{+5.1}_{-10.5}(exp)^{+5.1}_{-9.1}(f_0)^{+4.8}_{-6.4}(\omega_{B_s}))^\circ
 \end{equation}
or
\begin{equation}\label{theta1}
 \theta=(145.97^{+10.5}_{-5.1}(exp)^{+9.1}_{-5.1}(f_0)^{+6.4}_{-4.8}(\omega_{B_s}))^\circ
 \end{equation}
 The first error is from  experimental error
of the branching ratio of $B_s^0\to J/\psi f_0(980)$, the second one is due to the error of  the
branching ratio of $f_0(980)\to\pi^+\pi^-$, the third one is induced
by
 the uncertainty of shape factor $\omega_{B_s}$ of $B_s$ meson wave function.
 There are also other  theoretical errors in our calculations, such
 as the uncertainty of final state meson wave functions and the
 known higher order corrections. Needless to see, the uncertainty of  obtained
 measurement through this method is large.

 In this direction, we can also use the experimental date of $B_s \to
J/\psi f_0(980)$ and Eq.~(\ref{brbstojpsif0},\ref{brbstojpsisigma})
to predict the branching ratio of $B_s \to J/\psi \sigma$,
\begin{equation}
    Br(B_s ^0\to J/\psi~\sigma)= (4.72^{+0.66}_{-0.62}(f_0)^{+0.62}_{-0.59}(\omega_{B_s})) \times 10^{-5}.
\end{equation}

 To determine
phase $ \beta_s$ in $B^0_s$ mixing accurately for probe of NP, it is necessary to determine $f_0-\sigma$ mixing angle more accurately. We  need  take the second method to determine the $f_0-\sigma$ mixing angle
$\theta$ with less uncertinties. From Eq.~(\ref{brbstojpsif0},\ref{brbstojpsisigma}), we can get the relation of the ratio of the branching ratios of $B_s^0\to
J/\Psi~f_0$ and $B_s^0\to J/\Psi~\sigma$ with the $f_0-\sigma$ mixing angle $\theta$,

\begin{equation}\label{rbrf0withsigma1}
    R_{f_0/\sigma}=\frac{Br(B_s^0\to J/\Psi f_0)}{Br(B_s^0\to J/\Psi~\sigma)}=\cot^2\theta
    \frac{(1-r_2^2+\frac{1}{2}r_{2}^4-r_{3(f_0}^2)}{(1-r_2^2+\frac{1}{2}r_{2}^4-r_{3(\sigma)}^2)}\;,
\end{equation}
 The mass of $f_0(\sigma)$ is far less than that of $B_s$ meson,  so $r_{3(f_0(\sigma))}^2$ is  negligible.

 The Eq.~(\ref{rbrf0withsigma1}) can be reduced into,
 \begin{equation}\label{rbrf0withsigma2}
     R_{f_0/\sigma}=\frac{Br(B_s^0\to J/\Psi f_0)}{Br(B_s^0\to J/\Psi~\sigma)}=\cot^2\theta \;,
\end{equation}

This means that the mixing angle $\theta$ can be extracted from the
ratio of the branching ratios of $B_s^0\to J/\psi~f_0(\sigma)$ with
negligible theoretical uncertainty. The uncertainty of $\theta$
determined in this method is mainly from the uncertainty of the
measured  ratio of the branching ratios of $B_s^0\to
J/\psi~f_0(\sigma)$.  In Fig.\ref{vrf0sigma} we show the variation
of the ratio of the branching ratios of $B_s^0\to
J/\psi~f_0(\sigma)$ with $\theta$. If the ratio of the branching
ratios of $B_s^0\to J/\psi~f_0$ to that of $B_s^0\to J/\psi~\sigma$
were measured, we could determine the mixing angle $\theta$ fairly
well. This is a good news to  determine phase $ \beta_s$ in $B^0_s$
mixing accurately for the probe of NP.

\section{Summary and Discussion} In this paper, we derive the decay amplitude of $B_s^0\to
J/\psi~f_0(\sigma)$
 and the relation of the branching ratios of
$B_s^0\to J/\psi~f_0(\sigma)$. We computed the factorizable
contributions in QCDF approach and the hard spectator scattering
diagrams in the perturbative QCD approach. The branching ratio of
$B_s^0\to J/\psi~f_0$  is in agreement with recent experimental
data. We also predict the branching ratio of $B_s^0\to
J/\psi~\sigma$ to be $(4.72^{+0.66}_{-0.62}(f_0)^{+0.62}_{-0.59}(\omega_{B_s})) \times 10^{-5}$.  We suggest two
methods to determine the mixing angle $\theta$ of $f_0$ and
$\sigma$. For the first method we get $f_0-\sigma$ mixing angle
$\theta$ to be about $(34.03^{+5.1}_{-10.5}(exp)^{+5.1}_{-9.1}(f_0)^{+4.8}_{-6.4}(\omega_{B_s}))^\circ$ or $(145.97^{+10.5}_{-5.1}(exp)^{+9.1}_{-5.1}(f_0)^{+6.4}_{-4.8}(\omega_{B_s}))^\circ$, which is
in consistency with others. The second method for determining the
mixing angle $\theta$ has little theoretical  uncertainty, but needs
the experimental data of both the branching ratio of $B_s^0\to
J/\psi~f_0(\sigma)$ as an input. We hope that the future experiment
will measure it.
\section*{Acknowledgement}

This work is supported  by the National Science Foundation of China under the grant No.10975077, 10735080, 11075168 and 11047014, and  by the Priority Academic Program Development of Jiangsu Higher Education Institutions (PAPD).

\section*{Appendix: Input Parameters And Wave Functions}
We use the following input parameters in the numerical calculations\cite{pdg2010,betas4,Gamiz:2009ku}
\beq
 \Lambda_{\overline{\mathrm{MS}}}^{(f=4)} &=& 250 {\rm MeV}, \quad
 \quad f_{B_s} = (0.231\pm 0.015){\rm GeV}, \quad  M_{B_s} = 5.366 {\rm GeV}, \non
 M_W &=& 80.41{\rm GeV},\quad
\tau_{B_s^0}=1.472\times 10^{-12}{\rm
 s},
 \label{para}
\eeq
   For the CKM matrix elements,  we adopt the wolfenstein
parametrization for the CKM matrix up to $\mathcal{O}$$(\lambda^
3)$\cite{pdg2010},
\begin{equation}
V_{CKM}= \left(           \begin{array}{ccc}
          1-\frac{\lambda^2}{2} & \lambda & A \lambda^3 (\rho-i \eta)\\
          -\lambda & 1-\frac{\lambda^2}{2} & A \lambda^2 \\
          A \lambda^3 (1-\rho-i \eta )&-A \lambda^2 & 1
          \end{array} \right) , \label{vckm}
\end{equation}
with the parameters $\lambda=0.2253, A=0.808, \bar{\rho}=0.132$ and
$\bar{\eta}=0.341$.

For the $B_s$ meson distribution amplitude, we adopt the
model\cite{bswavefunction}

\beq \phi_{B_s}(x,b) &=& N_{B_s} x^2(1-x)^2 \mathrm{exp} \left
 [ -\frac{M_{B_s}^2\ x^2}{2 \omega_{B_s}^2} -\frac{1}{2} (\omega_{B_s} b)^2\right],
 \label{eq:phibs}
\eeq

where $\omega_{B_s}$ is a free parameter and we take
$\omega_{B_s}=0.5\pm 0.05$ GeV in numerical calculations, and
$N_{B_s}=63.6708 $ GeV is the normalization factor for $\omega_{B_s}=0.5$ GeV.

The $J/\psi$ meson asymptotic distribution amplitudes are given by
\cite{BC04}
\begin{eqnarray}
\Psi^L(x)=9.58\frac{f_{J/\psi}}{2\sqrt{2N_c}}x(1-x)
\left[\frac{x(1-x)}{1-2.8x(1-x)}\right]^{0.7}\;,\nonumber\\
\Psi^t(x)=10.94\frac{f_{J/\psi}}{2\sqrt{2N_c}}(1-2x)^2
\left[\frac{x(1-x)}{1-2.8x(1-x)}\right]^{0.7}\;,\nonumber\\
\label{jda}
\end{eqnarray}
The wave function for $s\bar{s}$ components of $f_0(\sigma)$ meson are given
as in Ref.~\cite{chy0508104}


\end{document}